\documentclass{epl}

\usepackage{graphicx}

\title{Order and disorder in columnar joints}

\author{Lucas Goehring \and Stephen W. Morris}

\institute{ Department of Physics, University of Toronto, 60 St.
George Street, Toronto, Ontario, M5S 1A7, Canada}


\begin{document}
\maketitle

\begin{abstract}
Columnar joints are three-dimensional fracture networks that form
in cooling basalt and several other media.  The network organizes
itself into ordered, mostly hexagonal columns. The same pattern
can be observed on a smaller scale in desiccating starch. We show
how surface boundary conditions in the desiccation of starch
affect the formation of columnar joints.  Under constant drying
power conditions, we find a power law dependence of columnar
cross-sectional area with depth, while under constant drying rate
conditions this coarsening is eventually halted. Discontinuous
transitions in pattern scale can be observed under constant
external conditions, which may prompt a reinterpretation of
similar transitions found in basalt. Starch patterns are
statistically similar to those found in basalt, suggesting that
mature columnar jointing patterns contain inherent residual
disorder, but are statistically scale invariant.
\end{abstract}

\begin{figure}
\onefigure[width=5in]{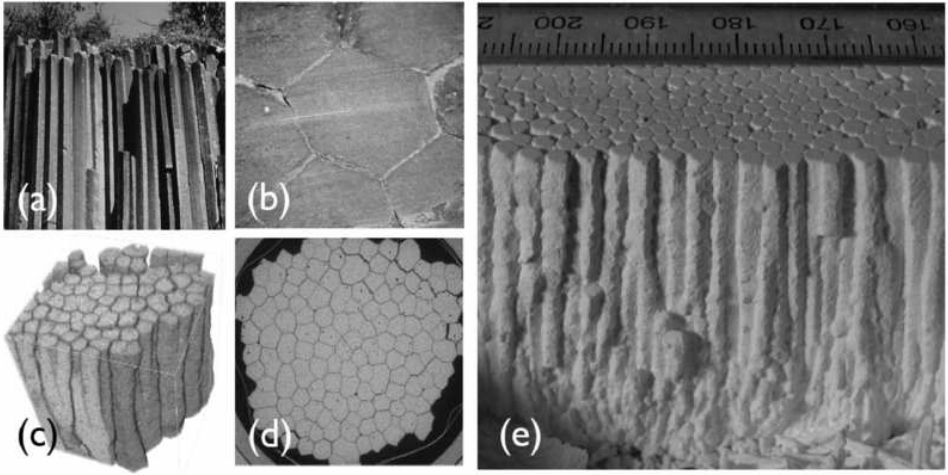} \caption{ Examples of
columnar jointing. (a) The colonnade of the Devil's Postpile, CA,
USA.  (b) Exposed surface of the Devil's Postpile, showing the
quasihexagonal fracture pattern, as it occurs in basalt. (c)
MicroCT x-ray tomography image of corn starch colonnade with 36
$\mu m^3$  voxel resolution, and ~2.5 cm/side. (d) Cross-section
of a tomogram at a depth of 18 mm. (e) A typical corn starch
colonnade (shown inverted) studied in these experiments, grown
with a constant evaporation rate.  } \label{examples_fig}
\end{figure}

Columnar jointing, in which shrinkage fractures arrange themselves
to leave behind a pattern of mainly hexagonal pillars, has
fascinated scientists and naturalists for centuries\cite{sirRBSRS,
orielly, huxley}.  The Giant's Causeway in Northern Ireland and
the Devil's Postpile in California are famous examples. Columnar
joints occur in lava flows, sandstone, mud, coal, glass, starch,
and ice\cite{huxley, french, degraff, muller1, muller2, muller3, Toramaru2004, seshadri, menger}, while
a 2D variant may cause ice/sand-wedge polygons in permafrost on
Earth and Mars\cite{degraff,jaglaPRE}.  Columnar joints are formed
through the self-organization of shrinkage cracks in a
three-dimensional material, as either heat or moisture is removed
from one surface\cite{muller1,budkew,peck}.  Brittle fracture is
initiated by the stress buildup at this surface\cite{peck}, and
the resulting cracks progressively extend, tracking a shrinkage
front as it moves into the bulk\cite{aydin,dufresne}.  At any
given time, active fractures are confined to a nearly planar layer
between the remaining compliant material, and the fully fractured
material\cite{aydin,dufresne}.  The evolution of this quasi-2D
polygonal pattern through time is recorded in the depth dependence
of the resulting prismatic columns.  Often, in geophysical
examples, this pattern only becomes apparent when the interior of
the formation is exposed by erosion and weathering.  As a result,
previous geophysical studies of the fracture pattern have been limited to
exposed planar surfaces.  Columnar jointing is a surprisingly
general phenomenon, well known in igneous rocks (both
terrestrial\cite{budkew} and lunar\cite{jones}), but also seen in
sedimentary\cite{seshadri, degraff}, and metamorphic
rocks\cite{degraff}, as well as in man-made and biological
materials\cite{degraff,french,muller1,muller2,muller3, Toramaru2004}. Examples in basalt and
corn starch are shown in Fig.~\ref{examples_fig}.  Joints can
range in size from micron sized diffusively cooled columns in
vitrified, impure, ice\cite{menger}  to meter sized evaporatively
cooled columns in basalt\cite{peck}.

Other polygonal fracture patterns confined to a plane may evolve
and order by similar processes, if a mechanism for the lateral
motion of cracks exists. For example, fractured freeze-thaw
polygons, with mobile edges, are observed in Arctic and Antarctic
permafrost\cite{lachen,sletten}. Similar features have been imaged
on Mars, suggesting the presence of Martian
permafrost\cite{sletten,mellon}.  Recent phase field models of
columnar jointing have been constructed to apply equally to all
these cases\cite{jaglaPRE}.  This generality suggests that a
detailed description of columnar jointing, based on controlled,
repeatable experiments, could have applicability to a number of
fields.

\begin{figure}
\onefigure[width=2.3in]{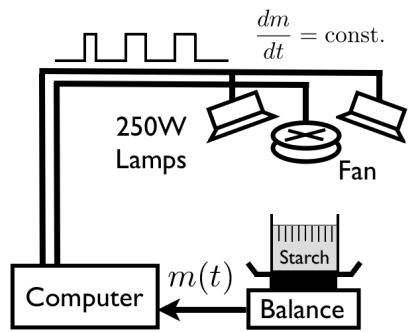} \caption{ Experimental
methods. An automated scale was used to feedback control the rate
of evaporation. After drying, the fracture pattern is measured
using a combination of MicroCT x-ray tomography and destructive
sampling. } \label{expt_fig}
\end{figure}

To date, due to the constraints on geophysical observation, there
exists no quantitative description of how columnar joints order.
In the absence of such experimental or field data, a number of
theories have been proposed
\cite{jaglaPRE, budkew, mallet, jagla2, jagla3}. However, there is no
consensus on exactly how or why these joints order. Nor is it
understood how surface boundary conditions and material properties
contribute to the columnar structure and scale.  In this paper, we
present the first ever 3D data describing columnar jointing,
specifically describing the ordering and coarsening processes
observed near the free surface.

Our experimental techniques are based on those of
M{\"u}ller\cite{muller1, muller2, muller3}, who independently
rediscovered the jointing in starches previously noted by
Huxley\cite{huxley} and by French\cite{french}.  These techniques
have also recently been adopted by Toramaru and Matsumoto
\cite{Toramaru2004}, who investigated the relationship between
desiccation rate and the starch pattern at a fixed depth.

250 W heat lamps were used to dry 1:1 slurries of 100\% pure corn
starch and water. Traces of bleach were used to sterilize the
experiment. We studied samples 1-100 mm deep, dried in circular
flat-bottomed dishes; evaporation rates were between 10-40 ${\rm
mg/h~cm}^2$, and samples typically dried between 1 and 28 days.
Water content was measured by weighing the samples once per minute
on an automated scale, as shown in Fig.~\ref{expt_fig}.
Desiccation rates were set by regulating the heating and
ventilation applied to the top surface of the starch. In one case
overhead lamps at a fixed starch-lamp distance supplied a
continuous, constant drying power. The drying power could be
changed between experiments by varying the lamp-starch distance.
In our second setup we used the measured sample weight, $m(t)$, to
drive a feedback loop which controlled the duty cycle of the heat
lamps and a small fan. We used this feedback to keep the
evaporation rate, $dm/dt$, constant.  We shall refer to these
methods as constant power and constant rate conditions,
respectively.  This level of control has not been available in the
previous experiments of M{\"u}ller\cite{muller1,muller2,muller3} or Toramaru and
Matsumoto\cite{Toramaru2004}.

As a starch slurry dries, first generation cracks appear,
penetrating through the entire sample depth and breaking the
sample into large, disconnected polygons. Later, much finer
secondary cracks initiate at the top surface, and propagate into
the sample. The secondary crack front leaves behind columnar
joints within each large polygon. This two-step fracture ensures
that container size and shape do not affect the columnar joint
pattern.

\begin{figure}
\onefigure[width=4.5in]{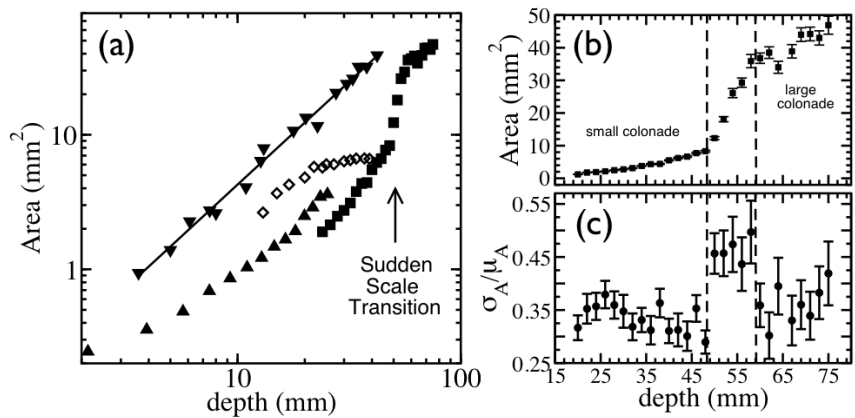} \caption{ The coarsening
behaviour of corn starch was studied in several drying situations.
(a) Generally, for samples with constant power (solid symbols), a
crude power law coarsening of cross-sectional area with depth was
seen, with exponents of 1.6 to 2.2 (the line shown has a slope of
1.6). As expected\cite{budkew}, faster drying rates produce
smaller columns.  The columns shown with inverted triangles were
grown with an initial drying rate of $\sim 10 {\rm mg/h~cm}^2$,
squares represent experiments with an initial drying rate of $\sim
32 {\rm mg/h~ cm}^2$. For samples dried with constant rate
conditions (open symbols), the coarsening is halted.  In cases of
deeper samples (squares), a sudden transition in scale was
observed, without an accompanying discontinuity in the drying
rate. (b) This transition (between dashed lines) is a sharp
discontinuity in scale, and is associated with increased pattern
disorder. (c) shows the relative disorder, as measured by the
standard deviation of the area distribution ($\sigma_A$), divided
by the mean of the area distribution ($\mu_A$). }
\label{coarsening_fig}
 \end{figure}

We have used MicroCT x-ray tomography, shown in
Fig.~\ref{examples_fig}c,d, to produce fully 3D visualizations of
the pattern of columnar jointing in desiccating starch under
several drying conditions. We have also recorded this pattern in
cross-section by measuring the counterparts left on the drying
containers after the starch is removed, and by sawing samples open
in order to destructively observe the pattern at different depths.
The violent nature of this destructive sampling constrains its use
to samples $> 1$~cm height, limiting data collection in some
cases.

\begin{figure}
\onefigure[width=4.2in]{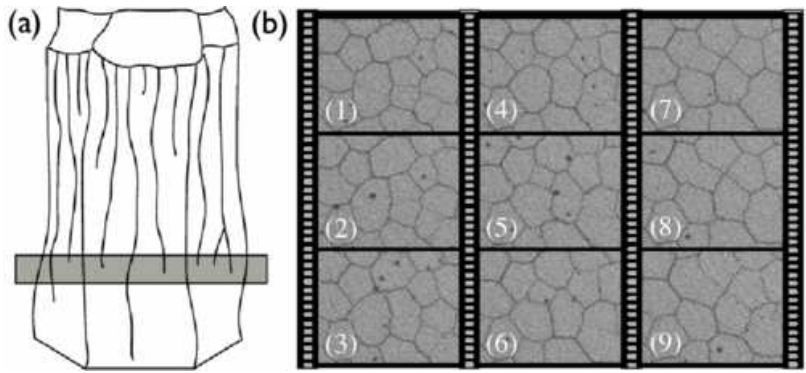} \caption{ Pattern
coarsening processes. (a) sketches the essential features of
coarsening through fracture termination, the motility of the crack
edges, and column creation from existing vertices. The upper
surface is covered with a fine network of fractures, which we have
omitted for simplicity.  The grey region demonstrates where a
burst of crack termination causes a sudden change in pattern
scale. (b1-9) show the pattern features in real space, and are
selected from a tomogram volume-filling image. Each panel shows a
4.5 x 3.5 mm$^2$ cross-section from approximately the centre of
the volume, at depths increasing by 360 $\mu$m per panel, 
beginning at a depth of 10.0~mm. The black
circles are bubbles, and the faint stripes are tomographic
artifacts. Note how mobile the column edges can be. Column merger
occurs between two pairs of columns near the centre of panel 1,
and between panels 3 and 4, in the lower right corner.  Either a
single crack, or a vertex can cease to propagate, and the merged
column rapidly readjusts its shape and cross-sectional area to
match its surroundings. Between panels 3 and 4, a new column
appears from a vertex near the centre of the image.}
\label{schematic}
\end{figure}

Fig.~\ref{coarsening_fig}a shows the evolution of the average
columnar cross-sectional area as a function of the depth.  Under
constant power conditions, the evaporation rate drops steadily, as
it becomes more difficult to drive water from the interior of the
sample. The rate of change of evaporation suggests that water
transport is partially driven by a wicking process, as is seen in
the drying of other suspensions\cite{dufresne}, and not, as
M{\"u}ller\cite{muller1} assumed, entirely by diffusion.  In fact, both diffusive heat transport and convection in the cracks are active in basalt\cite{hardee}. 
In our starch, water transport
is by wicking through the bulk of the material, rather than via the cracks.  These 
conditions result in a slowly decelerating drying front, and a
power-law coarsening of the columnar area.  This limited power law
breaks down in several suggestive ways.  The fracture scale is
limited by a fine surface crack pattern, which is well fit by
adding a small constant to the power law. This suggests that the
mature pattern scale is independent of the surface fracture scale.
In constant rate experiments, after an initial transient, the
drying front moves at constant speed, and the coarsening of the
pattern is almost entirely halted.   The transient coarsening is
indistinguishable from that observed using constant external
drying conditions.

In deep ($ > 5$ cm) samples, sudden transitions in scale were
observed to occur (see Fig.~\ref{coarsening_fig}b).  These
transitions are sharp, quite reproducible, and separate colonnades
of very different scales. Within the transition region, the
fracture pattern shows increased disorder, as shown in
Fig.~\ref{coarsening_fig}c.  The inspection of tomograms shows
that mergers of two or three columns, through the termination of a
common fracture or junction, are the only events that can lead to
pattern coarsening. Fig.~\ref{schematic} shows several mechanisms
of column evolution. New columns were occasionally created at
existing vertices, but no columns were seen to vanish by
constricting into a vertex. These observations imply that the
transition regions are zones of greatly enhanced merger rate. This
naturally increases the variance of the average column area during
the transition.

\begin{figure}
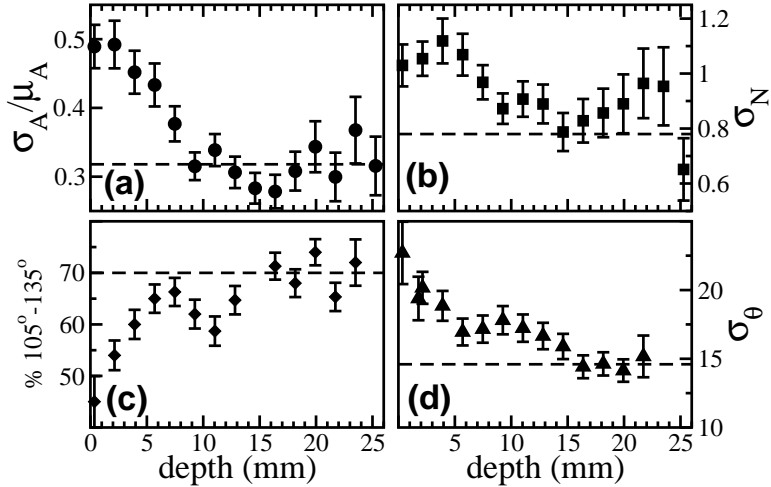

\onefigure[width=4in]{GM04_figure4.eps} \caption{ Statistical
comparison of a tomography study of a 28 mm deep corn starch
sample dried with constant power (data points, corresponding with
data shown in upright triangles in Fig.~\ref{coarsening_fig}a),
and O'Reilly's 1879 survey of the Giant's Causeway\cite{orielly}
(dashed lines). (a) The relative variation in column
cross-sectional area.  (b) Standard deviation of the distribution
of number of neighbours ($\sigma_N$). (c) \% of Y-joints, here
defined as those within $15^\circ$ of a $120^\circ$ joint. (d)
Standard deviation in the distribution of joint angles
($\sigma_\theta$), in degrees. } \label{stats_fig}
\end{figure}

Pattern disorder is increased during a sudden scale transition,
but some ordering process acts efficiently to return the value of
the relative variation in area to its pre-transition value of 0.35
within one data point (2 mm) of the end of such a transition.
Similar ordering occurs near the drying surface.  We investigated
this ordering behaviour in a 28 mm deep, continually coarsening,
starch colonnade, to study how a disordered surface crack network
matured. The plots in Fig.~\ref{stats_fig} show the depth
dependence of four statistics we used to quantify disorder.

We observe the evolution of an initially disordered surface
fracture pattern into a well-ordered colonnade.  To quantify this
ordering we have studied the distributions of areas, joint angles,
and the number of neighbours as a function of depth, measured from
the drying surface after complete desiccation, in cross-sections
of the tomogram.  Note that the number of neighbours, unlike the
number of sides, which is traditionally reported, must average 6
for a polygonal tiling of the plane that avoids
X-junctions\cite{gray}.

All four starch statistics reach a statistically steady state
after $\sim 1$ cm of evolution, maintaining fixed, but large
values thereafter.  This indicates that considerable disorder
remains in the pattern.  Furthermore, in all four cases, the
plateau values of the starch statistics match the values for the
Giant's Causeway\cite{orielly}, a mature, well-ordered basalt pattern\cite{budkew}.  
In this
experiment, it could be argued that the residual disorder is
dictated by the continued coarsening.  However, we observe
statistically similar descriptions of the limiting pattern in both
the constant rate experiments, for which coarsening is effectively
halted, and the exposed features of the non-coarsening basalt of
the Giant's Causeway.  The similarity in statistics from two very
different systems, desiccated starch and cooled basalt, combined
with evidence of a strong ordering processes away from the common
limiting pattern, suggests that residual disorder is intrinsic to
the quasihexagonal pattern of columnar jointing. This is contrary
to the frequently encountered assumption that columnar jointing
tends towards a perfect hexagonal lattice in
cross-section\cite{budkew, mallet}.  However, such an assumption
is based on the argument that hexagonal fracture maximizes the
elastic energy release\cite{mallet}. In strongly non-equilibrium
situations, such as this one, energy arguments are not necessarily
valid.

Our results can be used to explain a puzzling feature of basaltic
columnar joints.  A single basalt flow can contain several
adjacent colonnades, with different scales, separated by
disordered zones ({\it entablature}) often less than 1 m
wide\cite{gross, long}.  Some colonnade-colonnade boundaries occur
when cooling fronts travelling from the top and bottom of a flow
collide.  However, other transitions remain unexplained, except
through proposed catastrophic events such as intermittent
flooding\cite{long, degraff2}.   We suggest that entablature and
scale changes in basalt could occur even without sudden changes in
the external conditions.  In starch, we observed transitions in
scale using constant power external conditions, in which the
drying front is slowly, continuously, decelerating with depth.
Starch columns are approximately 100 times smaller in diameter
than basaltic columns. Directly scaling the 1 cm wide scale
transitions in starch, we find a 1 m scale that corresponds
reasonably well with the width of the entablatures observed in
basalt.  Such transitions could be the result of dynamical
instabilities of the fracture pattern, which occur when a stable
range of pattern scales is exceeded. An analogous instability
occurs in directionally dried thin films, in which a periodic
array of cracks can continue to propagate under a range of drying
conditions\cite{shorlin}. Outside this range, the array becomes
unstable, and makes a transition to a new spacing\cite{shorlin,
jagla2}.

Our dynamical instability-driven entablature hypothesis is
testable by observing the striae widths in basaltic colonnades.
Striae are 'chisel' marks on the sides of columns, which record
individual fracture advances as the cooling front
advances\cite{degraff}.  By observing the variation of the striae
width across a colonnade-colonnade scale transition it should be
possible to determine how the cooling conditions changed.  A
continuous evolution across such a boundary, rather than a
matching discontinuity in striae scale, would confirm the absence
of catastrophic changes in the external conditions.

Our experiments have focussed on exploring the 3D structure of
columnar jointing in corn starch.  We have directly observed the
operation of a strong ordering process that results in the eventual saturation of
the pattern statistics. This mature state of the pattern
contains considerable disorder, which we
suggest is due to the strongly nonequilibrium, complex dynamics of the joints.
An alternate explanation of residual
disorder is that the system's geometry 
evolves toward equilibrium but gets stuck in a local minimum
of the free energy \cite{jagla3}.  This scenario is difficult to completely rule out, but
seems unlikely given the observed persistent mobility of the joints,
which do not fluctuate about equilibrium positions. Several new
dynamical models of columnar jointing have recently been proposed
by Jagla\cite{jaglaPRE, jagla2}, and Jagla and Rojo\cite{jagla3},
some of which seem to capture, qualitatively, the disordered
dynamics we find.

We have observed coarsening of the starch colonnade as it
penetrates our samples. This coarsening proceeds through the
termination of crack tips along column sides or joints.  The
coarsening can be halted in constant rate experiments, revealing
an important relationship between fracture advance rate and
pattern scale. The observation of discontinuous transitions in
scale, however, imply it is not a one-to-one relationship. Rather,
as occurs with similar 2D patterns \cite{shorlin}, the exact
pattern scale will depend on the system's dynamical history, as
well as on the current externally imposed conditions. This alone
may significantly modify field interpretations of columnar
jointing, as we have outlined above.  However, the appreciation of
quasihexagonal fracture as a non-equilibrium dynamical system can
go much further, and a quantitative understanding of this pattern
could provide volcanology, cryophysics, planetary physics, and
pattern physics with novel diagnostic tools.  These considerations
motivate a continuing interest in this beautiful phenomenon.

\acknowledgments We thank Mark Henkelman and the Mouse Imaging
Centre for supplying access to micro-tomography equipment, and
Zhenquan Lin, A. Mark Jellinek, Pierre-Yves Robin, E. Alberto
Jagla, and R. Paul Young for helpful discussions.

  \end{document}